\documentstyle[12pt,epsfig]{article}
 \newcommand{\PP}{{\bf P}}
 \newcommand{\p}{{\bf p}}
 
 \newcommand{\0}{{\bf 0}}
 \newcommand{\x}{{\bf x}}
 \newcommand{\q}{{\bf q}}
 \newcommand{\s}{{\bf s}}
 
 \newcommand{\al}{\alpha}
 \newcommand{\poneos}{\stackrel{\not o}{\p\smash{\lefteqn{_1}}}}
 \newcommand{\psoneo}{\stackrel{o}{\p\smash{\lefteqn{'_1}}}}
 \newcommand{\poneo}{\stackrel{o}{\p\smash{\lefteqn{_1}}}}
 
 \newcommand{\ptwoo}{\stackrel{o}{\p\smash{\lefteqn{_2}}}}
 \newcommand{\ptwoos}{\stackrel{\not o}{\p\smash{\lefteqn{_2}}}}
 \newcommand{\pthros}{\stackrel{\not o}{\p\smash{\lefteqn{_3}}}}
 \newcommand{\pthro}{\stackrel{o}{\p\smash{\lefteqn{_3}}}}
 
 \newcommand{\po}{\stackrel{o}{\p}}
 \newcommand{\pso}{\stackrel{o}{\p\smash'}}
 \newcommand{\psoi}{{\stackrel{o}{\p}}'_i}
 \newcommand{\poi}{{\stackrel{o}{\p}}_i}
 \newcommand{\pok}{{\stackrel{o}{\p}}_k}

 \newcommand{\psio}{\stackrel{o}{\p\smash{\lefteqn{'_i}}}\phantom{_1}}

 \newcommand{\dvol}{\prod\limits_{k=1}^Nd\Omega_{\p'_k}d\Omega_{\p_k}}
 
 \newcommand{\ssumal}{\!\!\sum\limits_{\alpha',\alpha}}

 \newcommand{\dsty}{\displaystyle}

\begin{document}
\parindent=5mm
\normalsize

\begin{center}
{\bf\large Nucleon electromagnetic form factors \\
in a single-time constituent quark model}\\

\vspace*{3mm}
{\bf T.P.Ilichova, S.G Shulga}\\[6pt]
{\it Francisk Skaryna Gomel State University, \\
Laboratory of Particle Physics, JINR}  \\

\vspace*{3mm}
{\it Talk at the VIth International School-Seminar\\
      ``Actual Problems of Particle Physics''\\
    July 30 - August 8, 1999, Gomel, Belarus}
\end{center}

\begin{center}  {Abstract}  \end{center}
\footnotesize
\par
The main statement of the nucleon constituent quark model
with a fixed number of particles as basic ansatz
are considered
in a framework of the single-time (quasipotential)
approach to the bound state problem.
The scaling law breacking for the proton form factor 
are investigated for 
$Q^2 = 0 \div 2 $Gev$^2$.
\normalsize
\bigskip
\medskip
\par
The problem of the relativistic treatment of the constituent quark model
(CQM) have been
solved by several manners~\cite{CQM-ffactor1,CQM-ffactor2,CQM-ffactor3},
which have general feature - fixed number of
the particles assumption and additive assumption for nucleon current
expression in terms of quarks currents~\cite{Kobushkin}.
A popular approach to the problem
is the light-front dynamics~\cite{CQM-ffactor2}.
In present work we propose to use the old quasipotential approach~\cite{Logunov}
for formulation of CQM with fixed number of particles.
\par
In proceeding we suggest that Fock momentum-space for nucleon has $N=3$
quarks Dirac basis
$\mid\al_1\p_1,\al_2\p_2,\al_3\p_3\rangle
\equiv\mid\alpha {\bf p}\rangle$ :
\begin{equation}
\mid \lambda{\bf K}\rangle
\approx
\sum_{\alpha}\int
d\Omega_{\p_1}d\Omega_{\p_2}d\Omega_{\p_3} \mid\alpha {\bf p}\rangle
\langle\alpha {\bf p}
\mid \lambda{\bf K}\rangle.
\label{nonrel AKR}
\end{equation}
We will take it for granted that all
nontrivial effects of the QCD vacuum (e.g., gluon and quark-antiquark condensates)
can be absorbed into the effective parameters of the CQM.
\par
To find a equation for wave function (WF) $\langle\alpha {\bf p}
\mid \lambda{\bf K}\rangle$  we consider covariant single-time WF~\cite{Faustov}
($N=3$):
\begin{equation}
\begin{array}{l}
\Psi^{QP}_{\lambda{\bf K},\alpha}(x)=
 \Psi^{BS}_{\lambda{\bf K},\alpha}(x)
\delta(n_K(x_1-x_2))\ldots\delta(n_K(x_{N-1}-x_N)),\\
\Psi^{BS}_{\lambda{\bf K},\alpha}(x)=
\langle 0\mid T[
\phi^{(1)}_{\al_1}(x_1)\ldots\phi^{(N)}_{\al_N}(x_N)]
\mid \lambda{\bf K}\rangle,
\end{array}
\label{onetime wavef}
\end{equation}
where $\Psi^{BS}_{\lambda{\bf K},\alpha}(x)$ is Bethe-Salpeter WF;
$(n_K)_\mu=\frac{K_\mu}{\sqrt{K^2}}=\frac{K_\mu}{M_n}$ -
4-velocity,
$\delta$-functions in~(\ref{onetime wavef}) covariantly equal times
in the center-of-mass of system.
\par 
Using a translational invariance
$\phi(x_i)=\exp(i\hat H t_i)\phi(0,\x_i)\exp(-i\hat H t_i),$
 expression $\phi_{\al}(0,\x) =1/(2\pi)^3
\int d\Omega_q\sum_r\left[b_r(\q)u^{(r)}_\al(\q)\exp{(-iqx)}
+\ldots \right]$
and according to fixed number of particles assumption notice that
the anti-particle
operator do not give a contribution into the one-time WF
we can obtain the Fourier-momentum representation WF~(\ref{onetime wavef})
($D$ - the Wigner rotation matrix;$\po \equiv \poneo\;,\ptwoo\;,\pthro\;$,
$\poi = \vec L^{-1}_{K}p_i $)~\cite{Wigner,Faustov,Skachkov}:
\begin{equation}
\begin{array}{l}
\tilde \Psi_{\lambda\bf K,\alpha}(p)
=(2\pi)\delta^{(1)}(\stackrel{o}{P_0}-M_P)
\left[\sum_r
\prod_{i=1}^N u^{(r'_i)}_{\alpha_i}(\p_i)
\prod_{i=1}^N\stackrel{+}{D}\smash{^{r'_ir_i}_K}(p_i)\right]
\langle r\po\mid \lambda{\bf 0}\rangle
\label{psi_s_vig_pov}
\\
=(2\pi)\delta^{(1)}(\stackrel{o}{P_{0}}-M_P)
\prod_{i=1}^N u^{(r_i)}_{\alpha_i}({\bf p}_i)
\langle r\bf p\mid \lambda\bf K\rangle.
\end{array}
\label{psi_bez_vig_pov}
\end{equation}
\par
As in~\cite{Faustov}
we determine WF projection into positive-frequencial states:
\begin{eqnarray}
\tilde \Psi^{(+)(r)}_{\lambda{\bf K}}(p)\equiv
 \bar u^{r_1}_{\alpha_1}(\p_1)\ldots \bar u^{r_N}_{\alpha_N}
(\p_N) \tilde \Psi_{\lambda{\bf K},\alpha}(p)
=2\pi\delta(\stackrel{o}{P}_0-M_P)
\langle r{\bf p}\mid \lambda{\bf K}\rangle;
\label{psi_pol_chastotn}
\\
\langle r{\bf p}\mid \lambda{\bf K}\rangle=
(2\pi)^3\delta^{(3)}(\sum_{i=1}^N\stackrel{o}{\bf p}_i)
\prod_{i=1}^N\stackrel{+}{D}\smash{^{r'_ir_i}_K}(p_i)
\Phi^{(+)(r)}_{\lambda{\bf 0}}
(\stackrel{o}{\bf p}).
\label{analog_nerel_vyd_syscentermass}
\end{eqnarray}
\par
The connection established between CQM WF with fixed number of particle
$
\langle r {\bf\p}\mid \lambda\PP\rangle
$ and WF $\Phi^{(+)(r)}_{\lambda{\bf 0}}$
give us a chance to obtain equation for relative motion WF by standart form
~\cite{Logunov,Faustov}($\sum\po_i=\sum\psio=0$):
\begin{equation}
(\sum E_{\po_i}-M_P)
\Phi^{(+)(r)}_{\lambda\0}(\po)
=\int \prod_{i=2}^N d\Omega_{\psoi}
V^{(r'r)}(\po,\pso)
\Phi^{(+)(r')}_{\lambda\0}(\pso)
\label{uravnenie}
\end{equation}
\par
To construct a three-particle state with given total angular
momentum $J=1/2$ and helisity $\lambda$ we use the same method
as in Ref.~\cite{Terent}:
\begin{equation}
\Phi^{(+)(r)}_{\lambda\0}(\poneo\; ,\ptwoo\; ,\pthro\; ) =
\varphi_{\bf 0}^{model}{\chi}_\lambda^{SU(6)}(r_1,r'_2,r'_3)
D^{(r_2,r'_2)}_{(p_2+p_3)}(\ptwoos\;)
D^{(r_3,r'_3)}_{(p_2+p_3)}(\pthros\;),
\end{equation}
where $\poneos\; = - \ptwoos\ = {\vec L^{-1}_{p_1+p_2}}$.
We used the oscillator WF~\cite{Ilichova} and Coulomb-like WF for calculation
of the proton form factor:
\begin{equation}
\begin{array}{l}
\varphi_{\bf 0}^{osc} =
N\exp\left[ - \frac{2m}{3\gamma^2}
\left(
E_{\stackrel{o}{p}_1}+E_{\stackrel{o}{p}_2}+E_{\stackrel{o}{p}_3}-3m-
\sum\limits_{i\ne k}
\frac{(\poi\pok)}{\sqrt{(E_{\poi}+m)(E_{\pok}+m)
                       }
                 }
\right)
\right];
\\
\varphi_{\bf 0}^{Coulomb} =
N/(E_{\poneo}+E_{\ptwoo}+E_{\pthro})^2-9m^2+\gamma^2)^2.
\end{array}
\label{phi_cherez_exp}
\end{equation}
\par
Let us consider electromagnetic current matrix element.
Assuming that $J_{\mu}(0) = \sum\limits_{s=1}^{N}j_{\mu}^{(s)}(0) $
we first expand in sets of free-particle states ~(\ref{nonrel AKR}) and
current matrix element takes the form:
$$
     \langle n'\PP'|J_{\mu}(0)|n\0 \rangle =
      \sum\limits_{s=1}^N \int\dvol\ssumal
  \langle n'\PP'|\alpha'{\bf\p}'\rangle
  \langle\alpha'{\bf\p}|j_{\mu}^{(s)}(0)
         |\alpha{\bf\p}\rangle
    \langle\alpha{\bf\p}|n{\0}\rangle.
$$
For 3-quark system with symmetrical WF  we obtained
($\psoneo+\po_2+\po_3 = \p_1+\p_2+\p_3 = 0$):
 \begin{equation}
 \begin{array}{l}
 \langle n'\PP'|J_{\mu}(0)|n{\0}\rangle =
  3\int  d\Omega_{\p_2}d\Omega_{\p_3}
   \frac{\dsty m^2}{\dsty E_{\po_2+\po_3}
          E_{\stackrel{\phantom{0}}{\p_2+\p_3}}}
   \stackrel{\star}{\Phi}_{\lambda'0}^{(+)(r'_1,r'_2,r'_3)}
   (\psoneo\;,\po_2,\po_3)*
 \\{}*D\smash{^{r'_1\al'_1}_{P'}}(\p'_1)
    j_{\mu}^{\al'_1\al_1}(\p'_1,\p_1)
 D\smash{^{r'_2\al_2}_{P'}}(\p_2)
   D\smash{^{r'_3\al_3}_{P'}}(\p_3)
   \Phi_{\lambda 0}^{(+)(\al_1,\al_2,\al_3)}(\p_1,\p_2,\p_3)
 \end{array}
\label{protons current}
\end{equation}
\par
If $ \stackrel{o}{\s}=0$ then $ s=\frac{\stackrel{o}{s_0}}{M_P}P$ and for
$s=p'_1+p_2+p_3, s=p_1+p_2+p_3 $ we have
\begin{equation}
p\smash{'}_1+p_2+p_3=P'\frac{E_{\psoneo}+E_{\ptwoo}+E_{\pthro}}{M_P}\;;
p_1+p_2+p_3=P\frac{E_{\p_1}+E_{\p_2}+E_{\p_3}}{M_P}.
\end{equation}
Thus,~(\ref{protons current})
approximatly satisfies to current conservation condition:\\
if
$({p'}_1-p_1)_{\mu}
   \langle \al'_1\p\smash{^{'}_1}|j_{\mu}^{(1)}(0)|\al_1\p_1\rangle
=0
$ then
$(P'-P)_\mu
   \langle n'\PP'|J_{\mu}(0)|n{\0}\rangle \approx
$
$
\approx 0,\quad P\!\!=\!\!(M_n,\0).
$
The numerical calculation are shown that
this condition approximatly satisfy for wide region of the
model parameters. 6~-~dimensional integrals of this model
calculated by the Monte-Carlo method.
\par
\begin{figure}
  \epsfig{file=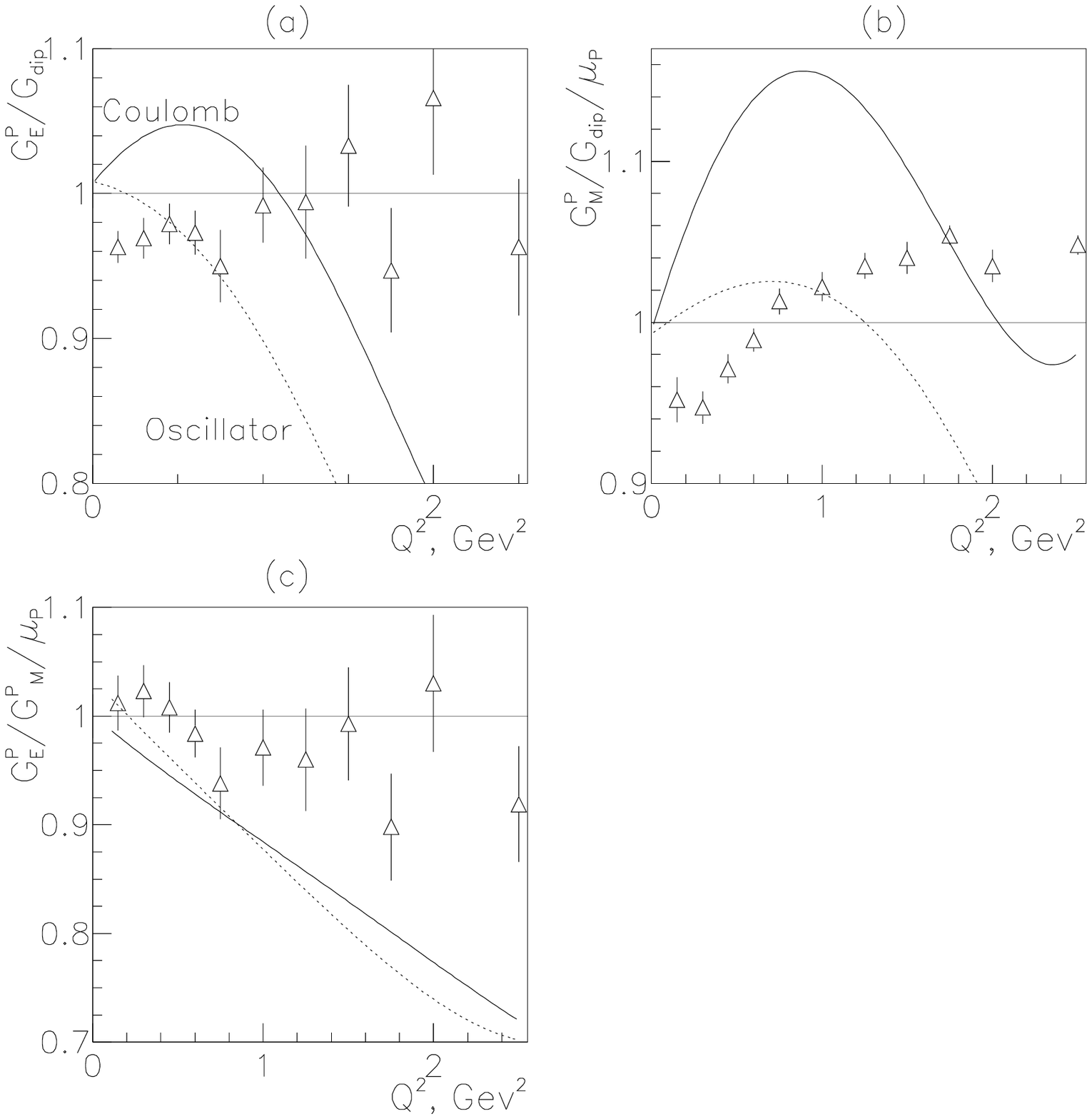,height=15cm}
	 \caption{  {\sl
 $G^P_E/G_{dip}$, $G^P_M/G_{dip}/\mu_P$  and $G^P_E/G_M/\mu_P$ in oscillator
 and Coulomb-like
 model with best parameters for fit :
Coulomb (solid curve): $m_q=166$ MeV, $\gamma^2=1.1$ GeV$^2$;
Oscillator (dashed curve): $m_q=162$ MeV, $\gamma^2=0.35$ GeV$^2$.
Experimental data
from~\cite{ffactor-exper}. $G_{dip}=1/(1+Q^2/0.71).$
	        }     }
\label{fig}
 \end{figure}

 Eq.~(\ref{protons current})
in nonrelativistic limit (additive quark model) is given by:
 \begin{eqnarray}
   \langle n'\PP'|J_{\mu}(0)|n{\0}\rangle
    \approx
3\stackrel{\star}{\chi}\smash{^{\al'_1,\al_2,\al_3}_{\lambda'}}\;
   \langle \al'_1\PP'|j_{\mu}^{(1)}(0)|\al_1{\bf 0}\rangle
   \chi^{\al_1,\al_2,\al_3}_{\lambda}
 F(t),
  \label{protons additive current}
 \end{eqnarray}
 where form factor is
$ F(t) =
 \int  d\Omega_{\p_2}d\Omega_{\p_3}
   \stackrel{\star}{\varphi}_{\bf 0}(\psoneo,\po_2,\po_3)
   \varphi_{\bf 0}(\p_1,\p_2,\p_3).$\\
Relation~(\ref{protons additive current}) provides the scaling law of the
space-like Sachs nucleon form factors 
($G^P_E \approx G^P_M/\mu_P$).
Dziembowski~\cite{CQM-ffactor2} obtained that the so-called soft
contributions (CQM) reproduce the experimental data extremely well 
up to the scale $Q^2\approx 2.5\; GeV^2$ for the nucleon.
Analogous results were obtaited by Cardarelli~\cite{CQM-ffactor2}
within light-front CQM without quark form factors.
We have reproduced it in our model (see fig.~\ref{fig},
experimental data
from~\cite{ffactor-exper}).
The Coulomb-like model describe best the experimental data up to 
$Q^2\approx 3\;$ GeV$^2$
for quark mass $m_q=166$ MeV, $\gamma^2 = 1.1$ GeV$^2$.
The magnetic moment ($\mu_P = G_M(0)$) was reproduced in this model
(with $\gamma^2 = 1.1$ GeV$^2$) for $m_q=105$ MeV but $G^P_E/G_{dip} \gg 1$,
 $G^P_M/G_{dip}/\mu_P \gg 1$ for $Q^2 = 0 \div 2$ GeV$^2$,
($G_{dip}=1/(1+Q^2/0.71$). 

\end{document}